\documentclass[conference]{IEEEtran}
\IEEEoverridecommandlockouts
\usepackage{cite}
\usepackage{amsmath,amssymb,amsfonts}
\usepackage{algorithmic}
\usepackage{graphicx}
\usepackage{textcomp}
\usepackage{xcolor}
\def\BibTeX{{\rm B\kern-.05em{\sc i\kern-.025em b}\kern-.08em
    T\kern-.1667em\lower.7ex\hbox{E}\kern-.125emX}}
\begin{document}

\title{An Incentive Security Model to Provide Fairness for Peer-to-Peer Networks
}

\author{\IEEEauthorblockN{1\textsuperscript{st} Samaneh Berenjian}
\IEEEauthorblockA{\textit{dept. Computer Science} \\
\textit{Stevens Institute of Technology}\\
New Jersey, USA\\
sberenji@stevens.edu}
\and
\IEEEauthorblockN{2\textsuperscript{nd} Saeed Hajizadeh}
\IEEEauthorblockA{\textit{dept. Mathematics, Statistics,}\\ \textit{and Computer Science} \\
\textit{University of Illinois at Chicago}\\
Chicago, USA\\
shajiz2@uic.edu}
\and
\IEEEauthorblockN{3\textsuperscript{rd} Reza Ebrahimi Atani}
\IEEEauthorblockA{\textit{dept. Computer Engineering} \\
\textit{University of Guilan}\\
Rasht, Iran \\
rebrahimi@guilan.ac.ir }
}

\maketitle

\begin{abstract}
Peer-to-Peer networks are designed to rely on resources of their own users. Therefore, resource management plays an important role in P2P protocols. Therefore, resource management  plays an important role in P2P protocols. Early P2P networks did not use proper mechanisms to manage fairness. However, after seeing difficulties and rise of freeloaders in networks like Gnutella, the importance of providing fairness for users have become apparent. In this paper, we propose an incentive based security model which leads to a network infrastructure that lightens the work of Seeders and makes Leechers to contribute more. This method is able to prevent betrayals in Leecher-to-Leecher transactions and more importantly, helps Seeders to be treated more fairly. This is what other incentive methods such as Bittorrent are incapable of doing. Additionally, by getting help from cryptography and combining it with our method, it is also possible to achieve secure channels, immune to spying, next to a fair network. The simulation results clearly show that how our proposed approach can overcome free-riding issue. In addition, our findings revealed that our approach is able to provide an appropriate level of fairness for the users and can decrease the download time.
\end{abstract}


\section{Introduction}
\label{sec:1}
A Peer-to-Peer (P2P) network is a distributed solution which uses the resources of its own users. In these networks, each node uses the resources of other users and also has the capability to provide resources for other nodes \cite{Tang2004,naghizadeh2017binary}. These networks have many applications, including file sharing, distributed processing and messages passing to other nodes which can cause many security problems \cite{naghizadeh2016structural,naghizadeh2015preserving}. P2P networks are highly dependent on resources of their own users \cite{Guo2016}. Therefore, the infrastructure of a P2P network should encourage users for contributions as well as using the resources of other members. The behavior of using resources of others without giving anything in return is called free-riding \cite{Hughes2005}. As a consequence, free-riders cause slower download times for contributing peers. Thus, it is an essential need for a P2P network to guarantee fair bandwidth allocation,  where a peer receives bandwidth equal to what it contributes. Accordingly, the system will be able to guarantee a certain level of performance for contributing peers.

Fair bandwidth allocation in P2P networks is not an easy task that can be tackled simply. This is due to the nature of P2P networks in which there is no central entity that can control and arbitrate access to all resources, and to schedule fair allocation of bandwidth for a router. Moreover, in P2P networks there is no perception in advance regarding the amount of bandwidth resources available and peers cannot be relied upon to specify their own resources honestly \cite{Balouchzahi2016}. Although free-riding can be a deliberate act, many users may become freeloaders or infringe fairness for lack of a proper infrastructure \cite{Lu2016}.

Since Seeders \footnote{For simplicity, we call nodes with all packets, Seeder and nodes without all packets, Leecher. These terminologies are taken from BitTorrent protocol.} are in the network only for altruistic reasons, a fair network should be designed in a way to put most of the traffics on the shoulder of Leechers. In this way, instead of getting free downloads, more users are encouraged to contribute to the network. Also, by mitigating the work of Seeders, we encourage them to stay longer in the network. For this purpose, we propose an incentive method which provides a proper infrastructure for fairness in P2P networks. This enables us to protect both Leecher-to-Leecher and Leecher-to-Seeder transactions. In other words, we do not only focus on protecting Leechers against freeloaders, but our goal is to protect Seeders as well. Furthermore, we show that by getting help and combining the standard methodology in cryptography, i.e. RSA, we can secure our channels against line spying, next to providing a fair network.

The rest of the paper is organized as follows. In section ~\ref{sec:SoA} we review the existing work for providing fairness in P2P networks. Section ~\ref{sec:2} introduces our proposed approach and its respective components as an incentive approach to provide fairness in P2P networks. In section ~\ref{sec:3} we elaborate our proposed approach by improving it with proper use of cryptography. In this section, we show how it is possible to protect communication channels as well.  In section ~\ref{sec:4} we analyze the functionality of our proposed method through simulations. Finally, we conclude the paper in section ~\ref{sec:5}.

\section{Related Work}
\label{sec:SoA}

Incentive-based systems usually consider some form of reward to encourage users for more cooperation. BitTorrent is one of the first major attempts which used incentive in its protocol \cite{Cohen2003}. It proposed a tit-for-tat (TFT) mechanism in order to incentivize peers to contribute resources to the system and discourage free-riders. As an important benefit, TFT encourages peers for more contribution without the need for centralized infrastructures. However, one important challenge regarding BitTorrent is that the robustness of the system is questionable. This is due to the fact that many of the contributions for improving the performance are unnecessary and can be reallocated or refused while still improving performance for strategic users. As a result, there are always some peers who contribute more data to the system than others.

It was shown that a cetralized authority can improve the fairness in BitTorrent \cite{naghizadeh2015counter}. In \cite{Izhak-Ratzin} and \cite{Li2011}, the authors used BitTorrent infrastructure to build their own fairness models. The main idea is to incentivize seeder stay longer in the BitTorrent network. The authors argued that download bandwidth of a leecher is the sum of all its associated peers' upload bandwidth, so redistribution of its associated peers' upload bandwidth could manage the leecher's download bandwidth. Although these approaches can help those peers who stay longer in the network get better download speed, in some cases they increase the overall download completion time.

Next to BitTorrent and its derivatives, there are other incentive methods which take an independent approach such as \cite{Karakaya2008, Joung2012, Kim2014}. In \cite{Karakaya2008}, the authors mainly focused on locating free-riders and taking actions against them. They proposed a framework in which each peer monitors its neighbors, decides if they are free-riders, and takes appropriate actions. Moreover, their approach does not require any permanent identification of peers or security infrastructures for maintaining a global reputation system. The results revealed that their proposed framework is able to reduce the effects of free- riding and can, therefore, increase the performance of a P2P network. Although this is an interesting approach, it suffers from a low number of downloads by contributors which arise from false detection in determining free-riders. This will ultimately influence the performance. 

In \cite{Joung2012}, the authors have proposed a Gnutella-like unstructured P2P file sharing network that utilizes free-riders to index files. This method is aimed to assist route query messages to destination peers. The authors suggested modifying a peer's neighbor table construction so that a free-rider has a lot of non-freeloaders as its overlay neighbors, while a non-freeloader's neighbors are mostly free-riders. Moreover, each peer also keeps information about its neighbors' files so that when a peer receives a query, it can determine if its neighbor has the queried file, or needs to forward query messages to continue the search. The findings obtained from this approach showed that the proposed method is able to improve search efficiency. However, one important security threat regarding this approach is that the authors assumed that peers faithfully execute the client program so that they follow the protocol designed by the program to build their neighbor tables and to relay query messages. If an adversary hacks the program, then it will be able to change the protocol. In \cite{Kim2014}, the authors suggested the use if a new P2P file sharing scheme based on the repeated Public Good (PG) game model. They mostly focused on trust evaluation, repeated interactions, and iterative self-learning techniques. The idea is to induce all peers to share files as many as possible by incorporating a trust-based P2P mechanism. The simulation results showed that the proposed scheme is able to maintain system efficiency as high as possible respond to current network conditions for adaptive management.


Trust management systems are methods to establish word-of-mouth for P2P networks. This means that based on transactions between nodes, they evaluate a degree of trust for each node which is mostly used for establishing a fair network. Finding a malicious user from its neighbors is a selection problem. Methods such as fuzzy decision making \cite{razeghi2015novel,berenjian2016intelligent} and genetic-based algorithms \cite{hatamian2016cgc} can be used to select the free-riders from multiple criteria. A successful trust management system that uses this approach is \cite{naghizadeh2016c}. The \cite{Gonçalves2016, Kamvar2003} are other examples of trust management systems. EigenTrust is the name of an approach proposed In \cite{Kamvar2003}. This method uses an algorithm to decrease the number of downloads of inauthentic files in a P2P network that assigns each peer a unique global trust value, based on the peer's history of uploads. For this purpose, the authors proposed a distributed and secure method to compute global trust values, based on power iteration. These values are then used by peers to select the peers from whom they download. 

As a result, the network will be able to identify malicious peers and isolates them from the network. EigenTrust is able to decrease the number of inauthentic files on the network, even under a variety of conditions where malicious peers cooperate in an attempt to subvert the system. In \cite{Gonçalves2016}, the authors have introduced an approach to predict a peer's cooperation level, focusing on the cooperation induced by the P2P protocol rather than the cooperation that results from user behavior or bandwidth limitation. This method mainly focuses on live streaming applications. By investigating the correlation between a peer's cooperation level, the authors tried to show that three centrality metrics, namely out-degree, out-closeness, and betweenness, are positively correlated with the cooperation level. Based on this, they proposed a non-linear regression model to measure peer's out-degree in the recent past to predict its cooperation level value in the near future.

\section{The Proposed Method}
\label{sec:2}

In this section, we propose an incentive model which can be used in versatile infrastructures. In our presented model, we increase fairness for both Leechers and Seeders. At first, we define how two Leechers should exchange packets and further demonstrate the mechanism for Leecher-to-Seeder transactions. 
 
\subsection{Protecting Leecher-to-Leecher Transactions}
\label{sec:2.1}
In this procedure, we aim to force Leechers to upload as much as they download from others. To this end, we consider $K'$ and $K''$ as two separate keys which are required by all nodes in the network. We define $E(X)$, for encryption  and $\mathcal{D}(X)$ for decryption of data $X$. Each node in the network is represented as $n_i$, $i=0, 1, \dots , k$. If $n_i$ wants to encrypt data $X$, it should use $K'_{n_i}$. Therefore, we define function $crypt(E(X), K'_{n_i})$ for encryption of data $X$ with key $K'_{n_i}$. On the other hand, for $n_i$ to decrypt data $X$, it is always required to use $K''_{n_i}$, we have $crypt({\mathcal{D}(X), K''_{n_i}})$ for decryption of data $X$ with key $K''_{n_i}$. 

\begin{figure}[t]
\centering
\includegraphics[width=70mm]{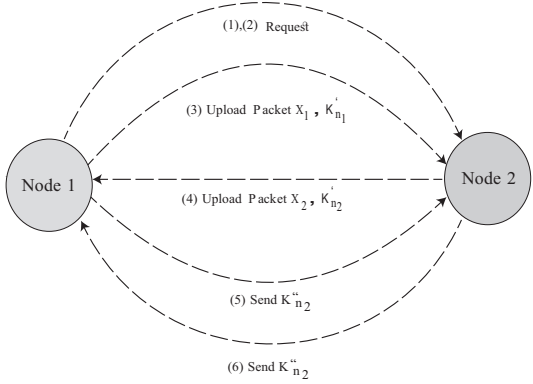}
\caption{The process of data exchange between two Leechers.}
\label{Fig1}
\end{figure}

Suppose $n_1$ and $n_2$ are two Leechers which want to exchange data $X_1$ and $X_2$. In the first step, $n_2$ sends the query package $Q_1 = (REQ, X_1, n_1)$ at which it requests data $X_1$ from $n_1$. After receiving $Q_1$, $n_1$ sends $Q_2 = (REQ, X_2, n_2)$  for $n_2$ to ask for data $X_2$. Before sending $X_1$, $n_1$ encrypts $X_1$ with $K'_{n_1}$. Accordingly, before sending $X_2$, $n_2$  encrypts $X_2$ with $K'_{n_2}$.

\begin{equation}
\label{eq1}
 \left\{
 \begin{array}{l  l} 
&n_2 \to n_1: Q_1 = (REQ, X_1, n_1)\\
&n_1 \to n_2: Q_2 = (REQ, X_2, n_2)\\
&n_1 \to n_1: crypt({E(X_1), K'_{n_1}})\\
&n_2 \to n_2: crypt({E(X_2), K'_{n_2}})
\end{array} \right.
\end{equation}

When the encryption process completes, package $Q_3 = (X_1, K'_{n_1}, n_2)$ will be sent to $n_2$. The decryption procedure for getting  $X_1$ requires $K''_{n_1}$, so $n_2$ must continue the transaction. After receiving $Q_3$ and performing the encryption process, $n_2$ will send the package $Q_4 = (X_2, K'_{n_2}, n_1)$ to $n_1$. To complete this procedure, nodes $n_1$ and $n_2$ need $K''_{n_1}$ and $K''_{n_2}$ respectively to decrypt their downloaded data. The process of exchanging the keys is done by $n_1$ which sends the package $Q_5 = (EXC, K''_{n_1}, n_2)$ to $n_2$. Doing this, it forwards  $K''_{n_1}$ to $n_2$. Accordingly, receiving $Q_5$, $n_2$ sends  $Q_6 = (EXC, K''_{n_2}, n_1)$ to $n_1$. As a result, $n_1$ gets $K''_{n_1}$.

\begin{equation}
\label{eq1}
 \left\{
 \begin{array}{l  l} 
&n_1 \to n_2:Q_3 = (X_1, K'_{n_1}, n_2)\\
&n_2 \to n_1: Q_4 = (X_2, K'_{n_2}, n_1)\\
&n_1 \to n_2: Q_5 = (EXC, K''_{n_1}, n_2)\\
&n_2 \to n_1: Q_6 = (EXC, K''_{n_2}, n_1)\\
\end{array} \right.
\end{equation}

At this stage, both nodes receive their requested data and have the keys for decryption. Using $K''_{n_2}$ and  $K''_{n_1}$, nodes $n_1$ and $n_2$ extract $X_2$ and $X_1$ respectively.

\begin{equation}
\label{eq1}
 \left\{
 \begin{array}{l  l} 
&n_1 \to n_1: crypt({\mathcal{D}(X_2), K''_{n_2}})\\
&n_2 \to n_2: crypt({\mathcal{D}(X_1), K''_{n_1}})\\

\end{array} \right.
\end{equation}

It is important to be pointed out that, afte receiving $Q5$, $n_2$ can leave the process without sharing $K''_{n_2}$ with $n_1$. Since we forced him to upload, economically we take away the incentives for betrayal. In real world applications, such behavior can even become disadvantageous. As nodes can mark them and refuse the future transactions. In Fig.~\ref{Fig1}, a better view of this process is demonstrated.

\subsection{Protecting Leecher-to-Seeder Transactions}
\label{sec:2.2}

To protect Seeders, we need to reduce their workloads by relegating the traffics to Leechers. In this way, we force Leechers to take responsibility for their downloads and engage them to upload process. As a result, Seeders by reducing their work are encouraged to stay longer in the network. This concept is first introduced in \cite{naghizadeh2016improving} with a P2P management system. In this work, we show that it is still possible to protect Seeders in an incentive model.

\begin{figure}[t]
\centering
 \includegraphics [width=60mm]{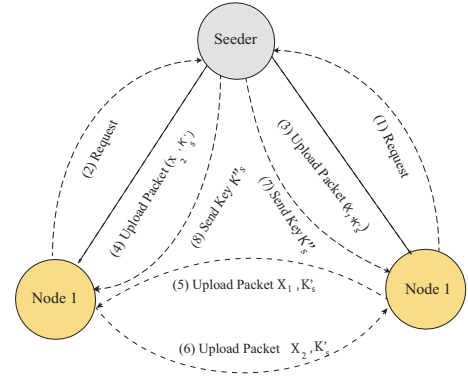}

\caption{The process of data exchange between two Leechers and a Seeder.}
\label{Fig2}
\end{figure}

Suppose that $n_1$ and $n_2$  are in contact with $s$, as the Seeder. Suppose $n_1$ sends $Q_1 = (REQ, X_1 \& X_2, s)$ to ask for $X_1$ and $X_2$ from $s$. In the same way, $n_2$ sends $Q_2 = (REQ, X_1 \& X_2, s)$ for asking the same data from $s$. After Seeder $s$ receives $Q_1$, it encrypt $X_1$ with $K'_{s}$. In the same way, after receiving $Q_2$, it  encrypst $X_2$ with $K'_{s}$. After encryption, $s$ sends $Q_3 = (X_1, K'_s, n_1)$ to send the encrypted data $X_1$ for $n_1$ and $Q_4 = (X_2, K'_s, n_2)$ to send the encrypted data $X_2$ for $n_2$.

\begin{equation}
\label{eq1}
 \left\{
 \begin{array}{l  l} 
&n_1 \to s: Q_1 = (REQ, X_1 \& X_2, s)\\
&n_2 \to s: Q_2 = (REQ, X_1 \& X_2, s)\\
&s \to s: crypt({E(X_1), K'_s})\\
&s \to s: crypt({E(X_2), K'_s})\\
&s \to n_1: Q_3 = (X_1, K'_s, n_1)\\
&s \to n_2: Q_4 = (X_2, K'_s, n_2)
\end{array} \right.
\end{equation}

For decryption process , both $n_1$ and $n_2$ need to obtain $K''_s$. In order to do this, $s$ can force them to cooperate and upload for each other. Therefore, to obtain $K''_s$, $n_1$ sends $Q_5 = (X_1, K'_s, n_2)$ to upload $X_1$ for $n_2$. Accordingly, $n_2$ sends $Q_6 = (X_2, K'_s, n_1)$ to upload $X_2$ for $n_1$. In this stage, both $n_1$ and $n_2$ shared their resources. For $n_1$ to obtain $K''_s$, $n_2$ sends $Q_7 = (ACK, X_1, s)$ to indicate $X_1$ is received and $s$ is allowed to share $K''_s$ with $n_1$. Node $s$ after receiving $Q_7$, sends $Q_8 = (EXC, K''_s, n_1)$ to the key $K''_s$ for $n_1$.

\begin{equation}
\label{eq1}
 \left\{
 \begin{array}{l  l} 
&n_1 \to n_2: Q_5 = (X_1, K'_s, n_2)\\
&n_2 \to n_1: Q_6 = (X_2, K'_s, n_1)\\
&n_2 \to s: Q_7 = (ACK, X_1, s)\\
&s \to n_1: Q_8 = (EXC, K''_s, n_1)\\
&n_1 \to s: Q_9 = (ACK, X_2, s)\\
&s \to n_2: Q_{10} = (EXC, K''_s, n_2)
\end{array} \right.
\end{equation}
For $n_2$ to receive the key, $n_1$ should apply for the same process. This means that after receiving $Q_8$, $n_1$ sends $Q_9 = (ACK, X_2, s)$ to indicate it has received $X_2$ from $n_2$. After receiving $Q_9$, node $s$ sends $Q_{10} = (EXC, K''_s, n_2)$ in order to forward $K''_s$ to $n_2$. In the last step, $n_1$ and $n_2$ which acquired $X_1$, $X_2$ and their keys,  decrypt $X_1$ and $X_2$. 

\begin{equation}
\label{eq1}
 \left\{
 \begin{array}{l  l} 
&n_1 \to n_1: crypt({\mathcal{D}(X_1), K''_s})\\
&n_1 \to n_1: crypt({\mathcal{D}(X_2), K''_s})\\
&n_2 \to n_2: crypt({\mathcal{D}(X_1), K''_s})\\
&n_2 \to n_2: crypt({\mathcal{D}(X_2), K''_s})
\end{array} \right.
\end{equation}
This protocol follows the procedure shown in equaton \ref{eq1} and the whole process is illustrated in Fig.~\ref{Fig2}.
 
\section{Improving the Basic Design}
\label{sec:3}

We can use the common protocols in cryptography to further improve our basic model in the previous section. In practice, by using symmetric key encryption, only one key is required for doing both process of encryption and decryption. As a result, we do not need to have two separate keys for each node. Also, we can further improve this method  and protect communications against line spying. To this end, we use RSA algorithm \cite{Rivest1978} which requires two new keys for each node. Suppose Alice ($A$) and Bob ($B$) want to exchange data $X_a$, $X_b$. First, $B$ sends the query package $Q_1 = (REQ, X_a, A)$ to request $X_a$ from $A$. After receiving $Q_1$, $A$ generates three different keys, $K_{a_{sy}}$, $K_{a_{pu}}$ and $K_{a_{pr}}$.

\begin{figure}[htp]
\centering
\includegraphics[width=70mm]{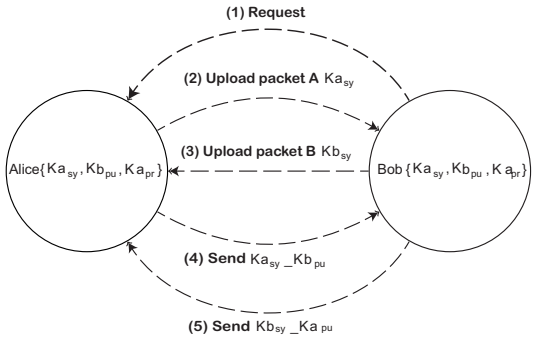}
\caption{The process of data exchange, using cryptography methods.}
\label{Fig3}
\end{figure}
The key $K_{a_{sy}}$ is the symmetric key which can be used to encrypt and decrypt data $X$. Keys $K_{a_{pu}}$  and $K_{a_{pr}}$ are used as a public and private key respectively. They are essential parameters in the $RSA$ algorithm. After generating the keys, $A$ sends query package $Q_2 = (EXC, K_{a_{pu}}, B)$ to exchange its public key with $B$. After $B$ receives the package $Q_2$, it follows the same pattern by generating $K_{b_{sy}}$ as its symmetric key, $K_{b_{pu}}$ as the public key and $K_{b_{pr}}$ as private key of $RSA$ algorithm. As like as the previous procedure, $K_{b_{sy}}$ is for encrypting and decrypting data $X$. After the keys are generated,, $B$ sends $Q_3 = (EXC, K_{b_{pu}}, A)$ to complete the public key exchange with $A$.

\begin{equation}
\label{eq1}
 \left\{
 \begin{array}{l  l} 
&B \to A: Q_1 = (REQ, X_a, A)\\
&A \to B: Q_2 = (EXC, K_{a_{pu}}, B)\\
&B \to A: Q_3 = (EXC, K_{b_{pu}}, A)
\end{array} \right.
\end{equation}

Once the process of exchanging $K_{a_{pu}}$ and $K_{b_{pu}}$ finishes, $A$ encrypts data $X_a$ with $K_{a_{sy}}$. When encryption process completes, $A$ sends the query package $Q_4 = (X_a, K'_{a_{sy}}, B)$ to $B$ and forward the encrypted data $X_a$ to it. After receiving $Q_4$, $B$  encrypts data $X_b$ with $K_{b_{sy}}$. Once this process completes, $B$ sends $Q_5 = (X_b, K_{b_{sy}}, A)$ to $A$ in order to deliver encrypted data $X_b$ to it.

\begin{equation}
\label{eq1}
 \left\{
 \begin{array}{l  l} 
&A \to A: crypt({E(X_a), K_{a_{sy}}})\\
&A \to B: Q_4 = (X_a, K'_{a_{sy}}, B)\\
&B \to B: crypt({E(X_b), K_{b_{sy}}})\\
&B \to A: Q_5 = (X_b, K_{b_{sy}}, A) 
\end{array} \right.
\end{equation}

In this stage, both nodes received their data. To exchange  $K_{a_{sy}}$ and  $K_{b_{sy}}$, we use $RSA$ methodology to be able to protect them from line spying. To exchange  $K_{a_{sy}}$ and $K_{b_{sy}}$, at first $A$  uses $crypt({E(K_{a_{sy}}}), K_{b_{pu}})$ to encrypt $K_{a_{sy}}$ with $K_{b_{pu}}$ and uses $Q_6 = (K_{a_{sy}}, K_{b_{pu}}, B)$ to send the encrypted key $K_{a_{sy}}$ for $B$. After receiving $Q_6$, $B$ uses $crypt({\mathcal{D}(K_{a_{sy}}), K_{b_{pr}}})$ to open the $K_{a_{sy}}$ first, and then uses $crypt({\mathcal{D}(X_a), K_{a_{sy}}})$ to receive data $X_a$. To complete the transaction, $B$ sends $Q_7 = (K_{b_{sy}}, K_{a_{pu}}, A)$ to send the encrypted key $K_{b_{sy}}$ for $A$. As the last step, $A$ at first uses $crypt({\mathcal{D}(K_{b_{sy}}), K_{a_{pr}}})$ to open the $K_{b_{sy}}$, and then uses $crypt({\mathcal{D}(X_b), K_{b_{sy}}})$ to receive data $X_b$.

\begin{equation}
\label{eq1}
 \left\{
 \begin{array}{l  l} 
&A \to A: crypt({E(K_{a_{sy}}}), K_{b_{pu}})\\
&A \to B: Q_6 = (K_{a_{sy}}, K_{b_{pu}}, B)\\
&B \to B: crypt({\mathcal{D}(K_{a_{sy}}), K_{b_{pr}}})\\
&B \to B: crypt({\mathcal{D}(X_a), K_{a_{sy}}})\\
&B \to A: Q_7 = (K_{b_{sy}}, K_{a_{pu}}, A)\\
&A \to A: crypt({\mathcal{D}(K_{b_{sy}}), K_{a_{pr}}})\\
&A \to A: crypt({\mathcal{D}(X_b), K_{b_{sy}}})
\end{array} \right.
\end{equation}

The above protocol is shown in Fig.~\ref{Fig3}. All data exchanges for both Leecher-to-Leecher and Leecher-to-Seeder, described in previous section should adapt and follow these steps. In this way, next to providing fairness for the network we can  shield the communication channels as well.

\section{Analysis and simulation results}
\label{sec:4}
In this section, we evaluate the effectiveness and functionality of our proposed approach through simulation. For this purpose, we use a network similar to Bitorrent protocol. In this environment, we initially compare our method with a network which only relies on willingness on the users. In order to appropriately examine the effectiveness of our approach, we mainly focus on the evaluation propagation, fairness for Seeders, fairness for all nodes (Seeders and Leechers) and free-riding prevention.     In the next step (see Section ~\ref{subsec:4.4}), by introducing tit-for-tat for nodes, we compare our method with Bittorrent.

\subsection{Propagation of Packets}
\label{subsec:4.1}    

Initially, we examine the packet propagation among nodes in the overlay. For this purpose, at each time interval, we calculated the average of all packets in the network. This shows how much time does it take for the nodes to complete their downloads tasks.

\begin{figure}[h]
\centering
\includegraphics[width=80mm]{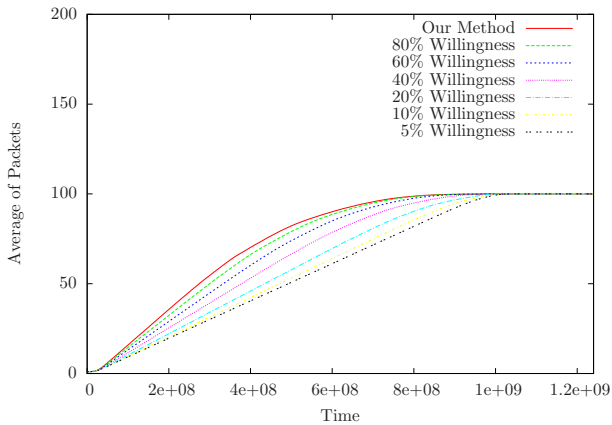}
\caption{Simulation results  show that using our method is beneficial for data propagation in the network.}
\label{Fig4}
\end{figure} 

For this experiment, we used a static network of 1000 nodes and initialised it with 10 Seeders. The data were divided into 100 smaller packets. Simulation stopped when all nodes in the network completed their downloads. The result is shown in Fig.~\ref{Fig4}.  As it can be seen, we compared our method with the willingness of users. By forcing users to upload more, they used the extra resources and could complete their downloads sooner. The results clearly shows that our approach enables nodes to download more packets in a certain time period compared to other willingness categories.

\subsection{Fairness for Users}
\label{subsec:4.2}

\begin{figure}[h]
\centering
\includegraphics[width=80mm]{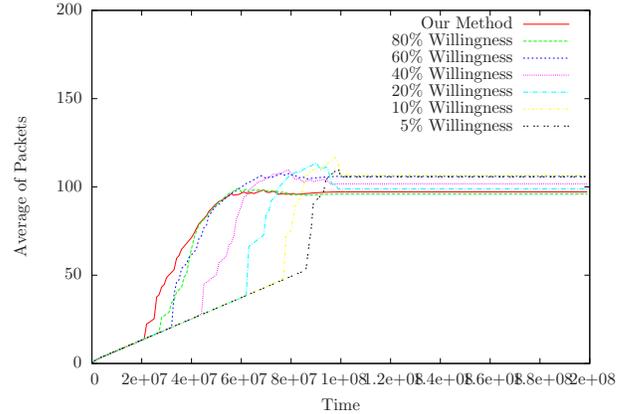}
\caption{Results of fairness Examination for Seeders.}
\label{Fig5a}
\end{figure}

To measure fairness for Seeders, the most important factor is their number of uploads. Therefore, we examine the mean of all uploaded packets for them at each time interval. For this experiment, a dynamic network of 1000 nodes is used. In this scenario, we assumed that Leechers would simply leave the network after completing their downloads. We used 10 Seeders and distributed them in the network evenly. Moreover, the data were divided into 100 smaller packets. Simulation was stopped when all nodes finished their downloads and left the network. The result of this experiment can be seen in Fig.~\ref{Fig5a}. 

Even though our method showed better performance compared to its rivals (except in one case), but the difference is not significant. This is mainly because we used only 10 Seeders against 990 Leechers. As a result, most of the transactions were between Leechers. It is expected when Seeders have more opportunities to relegate their uploads, the result will be more transparent. We further examine this situation in Section ~\ref{subsec:4.4} when we put our method against Bittorrent.

The overall average of uploads is useful to see how traffic is relegated from Seeders. But using overall uploads does not reveal fairness. To better understand the situation consider two pairs of nodes. One pair with (90, 10) uploads and the other pair with (50, 50) uploads. Even though both of them have the same mean, in the first pair, fairness is not served. Therefore, we have to use the variance of uploads. Variance shows to which extent nodes deviated their uploads from the mean of all uploads.

\begin{figure}
\centering
\includegraphics[width=80mm]{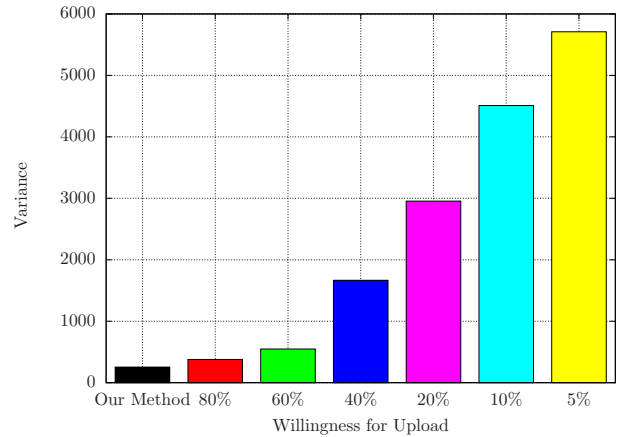}
\caption{Results of fairness examination for all nodes.}
\label{Fig5b}
\end{figure}

For this experiment we used a network of 100 nodes and 10 Seeders with 100 packets. We made sure that network is completely static so we could examine the behaviors of every node in the network. Simulation was running until all Leechers finished their downloads. As we can see in Fig.~\ref{Fig5b}, when less users are willing to share, most probably other users (mostly Seeders) get abused. This shows that when network does not implement a proper fairness mechanism, altruistic users are most vulnerable entities which suffer from such infrastructure. The result shows that as much as the willingness level decreases, the variance increases which is an indicator about the deviation from the mean of all uploads. In this experiment, our approach showed the minimum variance among other examined methods.

\subsection{Freer-riding Prevention}
\label{subsec:4.3}
Another important issue that we examine is free-riding prevention. The only way that freeloaders can live in a network shielded with our method is 1- using optimistic-unchock\footnote{In optimistic-unchoke, Leechers always uploads for a random node without considering if it is uploading back. In this way, new nodes also get a chance to acquire some packet. This helps newcomers to engage in future transactions\cite{Cohen}.} and 2- using Seeders when they do not have anyone to force them for contribution. Since the amount of downloads they can get from this way is significantly smaller, freeloaders have a hard time to finish their downloads.

\begin{figure}[h]
\centering
\includegraphics[width=80mm]{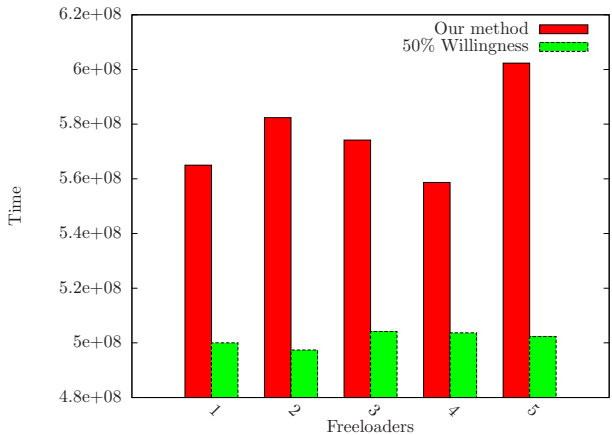}
\caption{The efficiency of our approach for keeping freeloaders for a much more longer time in the network.}
\label{Fig6}
\end{figure}	

For this experiment, we implemented a network with 500 nodes and 100 packets. We had 10 Seeders and 5 freeloaders which were distributed evenly in the network. Simulation was stopped when all freeloaders finished their downloads. We put our method against a 50\% willingness network. As it can be seen in Fig.~\ref{Fig6} with our method, freeloaders have to stay much longer in the network.  Clearly, almost in all scenarios, in the network which benefits from our approach, the freeloaders have to stay three times longer than a 50\% willingness network.

\subsection{A Comparison with Bittorrent}
\label{subsec:4.4} 
In a Leecher-to-Leecher transaction, unlike Bittorrent, we prevent betrayals before they take place. Another main advantage of our method comes from Leecher-to-Seeder transactions. The reason is that Bittorrent did not implement any mechanisms to help Seeders in the network. Therefore, we examine two main factors: 1- how much Leechers are forced to upload more and 2- how much Seeders upload less.

\begin{figure}[h]
\centering
\includegraphics[width=80mm]{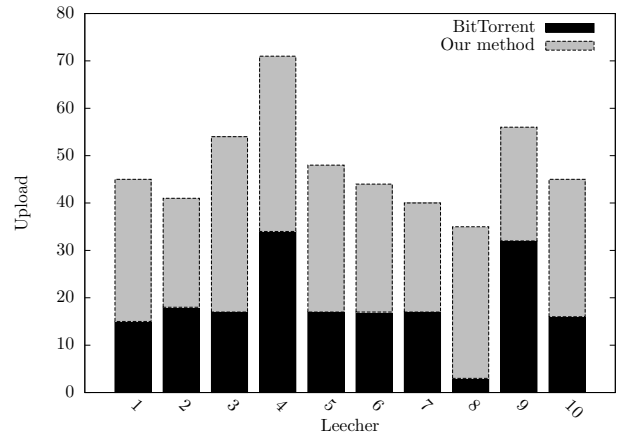}
\caption{The capability of our approach to encourage Leechers to upload more.}
\label{Fig7a}
\end{figure}	

Since Seeders play a crucial role in this experiment, we chose a smaller network with 30 nodes and 6 Seeders which were distributed evenly in the network. We used 100 packets in the network. In the first experiment (Fig.~\ref{Fig7a}), we show how much our method makes Leechers to take responsibility for their downloads. Simulation stopped when all of Leechers finished their downloads. After finishing downloads they could leave the network or stay and become another Seeder. For this purpose, 10 Leechers were chosen randomly for our analysis. The result is shown in  Fig.~\ref{Fig7a}. As we can see, in most cases, by using our method, Leechers uploaded more. In addition, compared to Bittorent, the reliance on Seeders is lower. 

\begin{figure}[h]
\centering
\includegraphics[width=80mm]{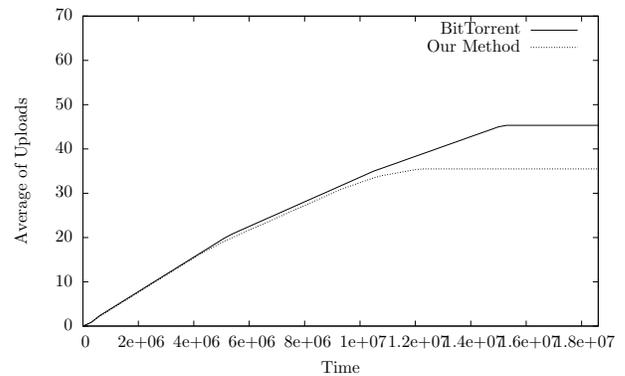}
\caption{With the proposed method Seeders tend to upload less.}
\label{Fig7b}
\end{figure}	

In the second experiment (Fig.~\ref{Fig7b}), we aimed to analyse the number of uploads for Seeders.  As we already mentioned, we had six Seeders in the network which were distributed evenly in the overlay. The average of uploads for all six Seeders at each time interval is shown in Fig.~\ref{Fig7b}. As it can be seen, as the time increases, the average of uploads of Seeders increases as well. However, in a network which uses our approach, this increment is slightly lower than Bittorent. This confirms the functionality of our approach against Bittorent. At the end of the simulation time, this difference in the average number of uploads, reaches its maximum limit. This reveals how our approach performs better as the time increases.

\section{Conclusion and Future Work}
\label{sec:5}
Fairness is one of the most important challenges in P2P communications.  In this paper, we proposed an incentive-based and fully distributed model that is aimed to accurately provide fairness in P2P networks. Our approach supports both Leechers and Seeders against freeloaders. The former is met by making Leechers upload as much as they download from others. The latter is guaranteed by reducing the workload of Seeders by relegating the traffics to Leechers. Consequently, we force Leechers to take responsibility for their downloads and engage in the uploading process. As a result, they are encouraged to stay longer in the network. We further showed that by combining the standard methodology in cryptography, i.e. RSA, we can secure our channels against line spying next to providing a fair network. Our findings obtained from the simulation results clearly shows the effectiveness and applicability of our approach by decreasing the download time, providing an appropriate level of fairness among peers, and preventing free-riding.




\end{document}